\documentclass{article}
\usepackage{graphicx,caption}
\usepackage{subfig}
\usepackage{float}
\usepackage{cite}
\usepackage{spconf,amsmath}
\usepackage{fontawesome}
\usepackage{IEEEtrantools}
\usepackage{amsmath}
\usepackage{amssymb}
\usepackage{tabularx}
\usepackage{soul} 

\usepackage{tikz}
\usepackage{tikz-3dplot}
\usetikzlibrary{calc,3d}
\usepackage{tikzsymbols}

\usepackage{epstopdf}
\usepackage{enumitem}

\usepackage{bm}
\usepackage{flushend}

\newcommand{\eM}{e_{q}^{(\mathrm{M})}}
\newcommand{\eV}{e_{v}^{(\mathrm{V})}}

\title{An Active Noise Control System Based On Soundfield Interpolation Using 
a Physics-informed Neural Network} 
\name{Yile (Angela) Zhang, Fei Ma, Thushara D. Abhayapala,  Prasanga N. Samarasinghe, and Amy Bastine
\address{Audio and Acoustic Signal Processing Group,  
The Australian National University, Australia
}
\thanks{Source code is at https://github.com/AngelaYZhang/pinn-assisted-anc. \\
This research is supported by an Australian Government Research Training Program (RTP) scholarship.
}}
\begin{document}
\maketitle
\begin{abstract}
Conventional multiple-point active noise control (ANC) systems require placing error microphones 
within the region of interest (ROI), inconveniencing users.
This paper designs a feasible monitoring microphone arrangement placed outside the ROI, 
providing a user with more freedom of movement.
The soundfield within the ROI is interpolated from the microphone signals using a 
physics-informed neural network (PINN).
PINN exploits the acoustic wave equation to assist soundfield interpolation under 
a limited number of monitoring microphones, and demonstrates better interpolation 
performance than the spherical harmonic method in simulations. 
An ANC system is designed to take advantage of the interpolated signal to reduce noise signal within the ROI.
The PINN-assisted ANC system reduces noise more than that of the multiple-point ANC system in simulations. 

\end{abstract}

\section{Introduction}
\label{sec:intro}
Active noise control (ANC) systems reduce unwanted primary noise by superimposing it with secondary noise generated by secondary sources~\cite{nelson1991active}. 
Multi-channel filtered-x least mean square (FxLMS)-based ANC system were proposed to reduce noise within a region of interest (ROI) using multiple reference and error microphones, and secondary sources \cite{elliott1987_multichannel,shi2019_multichannel}. 
Multiple-point ANC system is a type of multi-channel ANC system that reduces noise around error microphones. To reduce noise within the ROI, this system requires placing monitoring microphones around user's head to measure the residual noise field, which is undesirable.

To resolve this constraint, researchers proposed to use the remote microphone (RM) technique to estimate soundfield at ROI without any microphones inside the ROI~\cite{moreaureview,qiureview}.
This technique provides more freedom of movement to the user, making ANC system more practical~\cite{elliott_rm}.
Arikawa \textit{et al.}~\cite{tokyo_interpolation} used reference microphones placed outside of the ROI to interpolate the primary noise field and the residual noise field within the ROI. 
However, they assumed the number of reference microphones to be relatively large.
Jung \textit{et al.}~\cite{jung_rm} proposed an ANC headrest system with $16$ remote monitoring microphones placed around user's head and evaluated ANC performance on automobile road noise data.
These RM techniques interpolate the virtual microphone signals based on monitoring microphone signals 
only, and thus their performance is limited by number of monitoring microphones.

Recently, researchers exploited spatial soundfield characteristics to improve soundfield interpolation performance. When combined with the RM technique, they formulated feasible ANC system with improved system setup.
They proposed to decompose soundfield onto basis functions using spherical harmonic (SH)~\cite{june_rmSH,maeno_SH} and 
singular value decomposition~\cite{qiu_rm}, and interpolated soundfield with a 
small number of microphones. 
A drawback of these decomposition approach is the need to fully surround ROI with 
microphone arrays, either in spherical or multiple circular setup which still restrict user's access to the ROI.
Chen \textit{et al.}~\cite{hanchi} simplified the microphone array structure by designing a compact 2D microphone array for 3D soundfield estimation using SH. However, this design requires a large number of microphones and is infeasible in practice.
Maeno \textit{et al.}~\cite{maeno_SH} improved the ANC system setup by placing multiple compact reference microphone arrays close to the noise sources,
although the error microphones remain surrounding the ROI. 


Motivated by previous works, we design a practical microphone arrangement and propose a physics-informed neural network (PINN)-based interpolation method~\cite{xingyu_PINN,fei_pinn} for an ANC system.
In our proposed system, a small number of monitoring microphones are placed outside the ROI 
(around user's ears).
This is a more feasible arrangement compared to SH ANC system using spherical or circular arrays, providing user more movement flexibility. 
By integrating the monitoring microphone signal with the acoustic wave equation, 
we design a PINN that achieves more accurate soundfield interpolation 
than a SH-based interpolation method in the simulation.
Based on the interpolated soundfield, we use the multi-channel FxLMS algorithm to minimize 
the error signal and achieve better noise reduction within the ROI compared with the 
multiple-point ANC system.
The performance of this work is verified by simulations.

\section{Problem formulation}
\label{sec:problem}
\begin{figure}[t]
\centering
\includegraphics[width=7cm]{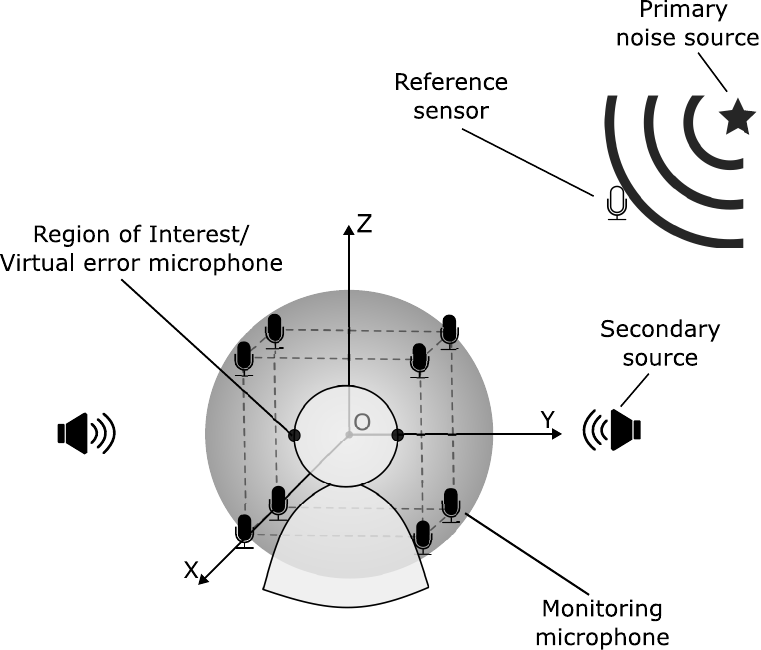}
\caption{System setup: A primary noise source generates the primary noise field and is detected by the reference sensor.
An ANC system cancels the primary noise field at the ROI (user's ears) by superimposing it with a secondary noise field produced by the secondary sources.
}
\label{fig:Mic setup}
\end{figure}


Consider an ANC system as shown in Fig.~\ref{fig:Mic setup} with $L$ secondary sources and $Q$ monitoring microphones.
A reference sensor is placed close to the primary noise source to detect the primary noise source characteristics, and the reference signal is denoted $x(n)$ with $n$ is the discrete time index.
Let $d_{\ell}(n)$, $\ell=1,\ldots,L$ be the secondary source signals and $\eM(n)$, $q=1,\ldots,Q$ be the received signal at the $q^\text{th}$ monitoring microphone located at $(\mathrm{x}_q,\mathrm{y}_q,\mathrm{z}_q)$ in Cartesian coordinates (or $(r_q,\theta_q,\phi_q)$ in spherical coordinates).
The received signal is
\begin{IEEEeqnarray}{rcl}
   \eM(n) = p_q(n) + \sum_{\ell=1}^{L} s_{\ell,q}(n) * d_{\ell}(n),
\end{IEEEeqnarray}
where $*$ is the convolution operation, $p_q(n)$ is the primary signal at the $q^{\text{th}}$ monitoring microphone, and $s_{\ell,q}(n)$ is the impulse response of secondary path from $\ell^{\text{th}}$ secondary source to $q^{\text{th}}$ monitoring microphone.

Consider $V$ virtual microphones positioned at or close to the two ears (ROI) at $(\mathrm{x}_v,\mathrm{y}_v,\mathrm{z}_v)$ with signal $\eV(n)$, $v=1,\ldots, V$. Although the virtual signals cannot be measured directly, they can be interpolated from the monitoring microphone measurements
\begin{IEEEeqnarray}{rcl}
    \eV(n) = \mathcal{I}(\eM(n)),
\end{IEEEeqnarray}
where $\mathcal{I}(\cdot)$ is the interpolation function.

The aim of our paper is twofold:
\begin{enumerate}[label=(\roman*)]
\item Interpolate 
 the virtual microphone signals $\eV$ based on the monitoring signals $\eM$ 
using a PINN.
\item Set up an ANC system to reduce the noise at ROI using FxLMS algorithm and the interpolated signals $\eV$.
\end{enumerate}

\section{Methodology} \label{sec:Methodology}

\begin{figure}[t]
\centering
\includegraphics[trim={6.3cm 2.2cm 7.7cm 3.1cm},clip,height=5.5cm]{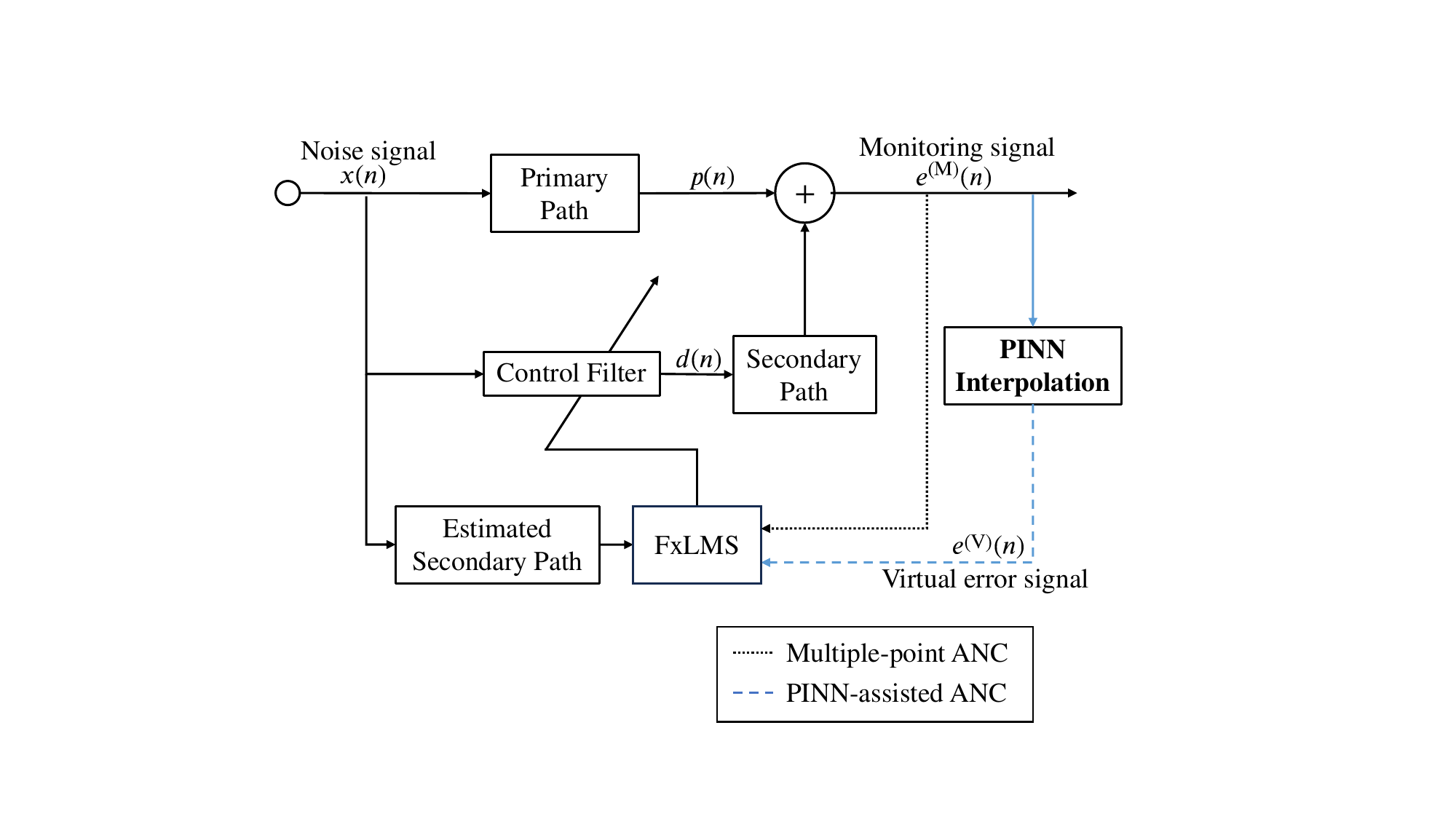}
\caption{Block diagram of multiple-point ANC system and PINN-assisted ANC system,
which differ by the error signal used in the FxLMS algorithm.}
\label{fig:fxlms_pinn}
\end{figure}

In this section, we first introduce the multiple-point ANC system, followed by 
formulating the PINN-assisted ANC system using PINN-interpolated soundfield. 
The two ANC systems are depicted in Fig.~\ref{fig:fxlms_pinn}.

\subsection{Multiple-point ANC System}
We consider the standard single-reference, multiple-output FxLMS algorithm~\cite{kuo_book}.
The weights of the adaptive filter for the $\ell^{\text{th}}$ secondary source is updated iteratively
\begin{IEEEeqnarray}{rcl}
\label{eq:weights_update}
    \mathbf{w}_{\ell}(n+1) = \mathbf{w}_{\ell}(n) + \mu \sum_{q=1}^{Q} \bm{x}_{\ell,q}^{\prime}(n) 
    \eM(n),
\end{IEEEeqnarray}
where $\mu$ is the step-size parameter,
and the filtered reference signal is
\begin{IEEEeqnarray}{rcl}
    \bm{x}_{\ell,q}^{\prime}(n) = s_{\ell,q}(n) * \bm{x}(n),
\end{IEEEeqnarray}
where $\bm{x}(n) = \left[x(n),x(n-1),\ldots,x(n-\mathcal{N}+1)\right]$ and $\mathcal{N}$ is the filter length.
The multiple-point ANC system aims to reduce noise $\eM(n)$ at multiple monitoring microphones.

\subsection{PINN-assisted ANC System}
\label{sec:pinn system}

Let the time variable $n$ and Cartesian coordinates $(\mathrm{x},\mathrm{y},\mathrm{z})$ be the input to a fully connected feed-forward network~\cite{raissi2017pinn},
which consists of one input layer, one output layer, and $\mathsf{L}$ hidden layers with $\mathsf{N}$ neurons in each hidden layer. 
The network output $\hat{p}(n,\mathrm{x},\mathrm{y},\mathrm{z})$ is the estimated primary signal at point $(\mathrm{x},\mathrm{y},\mathrm{z})$ at time $n$. 

We design the PINN to minimize a loss function 
\begin{IEEEeqnarray}{rcl}
&&\mathfrak{L}= 
\underbrace{
\frac{1}{Q}\sum_{q=1}^{Q} \left( \hat{p}(n_q,\mathrm{x}_q,\mathrm{y}_q,\mathrm{z}_q)-p(n_q,\mathrm{x}_q,\mathrm{y}_q,\mathrm{z}_q) \right)^2
}_{\mathfrak{L}_{\mathrm{data}}}
\nonumber\\
&&+
\underbrace{
\frac{1}{A}
\sum_{a=1}^{A}
\left( c^2 \nabla^2 {\hat{p}}(n_a,\mathrm{x}_a,\mathrm{y}_a,\mathrm{z}_a) 
-\frac{\partial^2}{\partial{n^2}} \hat{p}(n_a,\mathrm{x}_a,\mathrm{y}_a,\mathrm{z}_a) \right)^2
}_{\mathfrak{L}_{\mathrm{PDE}}},        \nonumber\\
\label{eq:lost_function}
\end{IEEEeqnarray}
where $\mathfrak{L}_{\mathrm{data}}$ represents the mean squared error loss between the PINN estimated primary signal 
and the ground truth primary signal $p(n,\mathrm{x},\mathrm{y},\mathrm{z})$ obtained at the monitoring microphone positions 
$\{\mathrm{x}_q,\mathrm{y}_q,\mathrm{z}_q\}_{q=1}^{Q}$, $\{\mathrm{x}_a,\mathrm{y}_a,\mathrm{z}_a\}_{a=1}^{A}$ is the Cartesian positions of $A$ randomly selected points around the ROI.
The PDE loss $\mathfrak{L}_{\mathrm{PDE}}$ is derived from a discrete approximation of the acoustic wave equation given by~\cite{fourier_book}
\begin{IEEEeqnarray}{rcl}
\nabla^2 p - \frac{1}{c^2} \frac{\partial^2 p}{\partial{}t^2} = 0,
\end{IEEEeqnarray}
where 
$ \nabla^2 \equiv \partial^2 / \partial{\mathrm{x}^2}
+ \partial^2 / \partial{\mathrm{y}^2}  
+ \partial^2 / \partial{\mathrm{z}}^2$ 
is the Laplacian operator, $\partial^2 / \partial{t^2} $
is the second partial derivative with respect to time $t$ and $c$ is the speed of sound.

The trained PINN model can interpolate the primary signal at the virtual microphones
and subsequently interpolate $\eV(n)$, $v=1,\ldots, V$. 
Our proposed PINN-assisted ANC system replaces $\eM(n)$ in~\eqref{eq:weights_update} of 
the multiple-point ANC system with $\eV(n)$, and aims to reduce noise $\eV(n)$ at the virtual microphone positions.

\section{Numerical experiments}\label{sec:exp}
\subsection{Experimental Settings}
We consider an experiment setup as in Fig.~\ref{fig:Mic setup} with $L=2$ and $V=2$. 
A single primary noise source is located at $(0.6, 0.8, 1)$ m and generates a tonal noise composed of three sinusoidal waves of 300, 400 and 500 Hz, each with a random phase. 
The speed of sound is set to $c=343 \mathrm{~m/s}$. We set the sampling rate as $24$ kHz and sample all signals for a duration of $0.1$ s.
Two secondary sources are placed at $(0, \pm 0.5, 0)$ m along the $y$-axis. 
Eight monitoring microphones $(Q=8)$ are placed on a $r=0.26$ m sphere at $(\pm 0.15, \pm 0.15, \pm 0.15)$ m, 
and the two virtual error microphones are at $(0, \pm 0.1, 0)$ m along the $y$-axis.
We model the primary and secondary paths using the free-field Green's function~\cite{fourier_book}.
This setting is applied to the PINN assisted-ANC system. 
The multiple-point ANC system cancels noise at the monitoring microphones without any interpolation.

\noindent\textbf{Interpolation using SH method:} 
The sound pressure of monitoring microphones at $(r=0.26 \; \mathrm{m},\theta_q,\phi_q)_{q=1}^{Q=8}$ can be decomposed onto the SH as~\cite{fei_anc}
\begin{IEEEeqnarray}{rcl}
\label{eq:SH_time}
p(n,r,\theta_q,\phi_q) \approx \sum_{u=0}^U \sum_{v=-u}^u \alpha_{u,v}(n,r) Y_{u}^{v}(\theta_q,\phi_q),
\end{IEEEeqnarray}
where $Y_{u}^{v}(\theta,\phi)$ is the SHs of order $u$ and degree $v$ with coefficient $\alpha_{u,v}$, $U = \lceil{2 \pi {f_\mathrm{m}} r / c}\rceil$~is the maximum order of soundfield~\cite{thushara_reproduction} with $f_\mathrm{m}$ being the highest frequency of interest.
The sound pressure at an arbitrary point $(r_s,\theta,\phi)$ can be interpolated as
\begin{IEEEeqnarray}{rcl}
\label{eq:SH_reconstruct}
p_s(n,r_s,\theta,\phi) 
&\approx& \sum_{u=0}^U \sum_{v=-u}^u 
\alpha_{u v}(n,r) \nonumber \\
&&\ast\mathcal{F}^{-1}\left[\frac{j_u(2\pi {f_\mathrm{m}}r_s/c)}{j_u(2\pi {f_\mathrm{m}}r/c)} \right]
Y_{u}^{v}(\theta,\phi),\quad 
\end{IEEEeqnarray}
where $\mathcal{F}^{-1}$ is the inverse Fourier transform and $j_u(\cdot)$ is the spherical Bessel function of the first kind~\cite{fourier_book}.

Accurate estimation of SH coefficients up to order $U$ requires the number of measurements $Q > (U+1)^2$~\cite{fei_pinn}.
With our setup $Q=8$, we set $U=2$ which is the closest fit to the accuracy criteria \cite{thushara_reproduction}.
We interpolate the soundfield on different spheres around the region for $r_s \in [0.1,0.4] $ m.
On each sphere, the soundfield is interpolated on 400 uniform sampling points~\cite{spherical_arrays_book}.
The sound pressure at each point is interpolated using~\eqref{eq:SH_time} and~\eqref{eq:SH_reconstruct}.

\noindent\textbf{Interpolation using PINN method:} 
We construct the PINN model to have $\mathsf{L}=1$ layer and $\mathsf{N}=16$ neurons. The activation function is $\tanh$.
Network parameters are initialized using the Glorot normal initializer~\cite{glorot} and use Adam~\cite{adam} as the optimizer.
We use the automatic differentiation from Tensorflow to calculate the partial derivatives of the pressure signal.
The network is trained for 5$\times$10$^5$ epochs with a learning rate of $0.001$. 
Time variable $n$ is normalized to the same range as the coordinate inputs, which is $[-0.15, 0.15]$.
The eight monitoring microphone positions are combined with 92 randomly selected positions within a $r=0.26$ m sphere to form $\{\mathrm{x}_a,\mathrm{y}_a,\mathrm{z}_a\}_{a=1}^{A=100}$. 
These positions and the time variable are inputs to the PINN model.

We compare the interpolation performance between the SH and PINN method by interpolating the pressure signal at the same positions.

\subsection{Results} \label{sec:subsec_results}
\begin{figure}[ht]
\centering
\includegraphics[trim={0.15cm 0cm 0.4cm 0.6cm},clip,height=4.4cm]{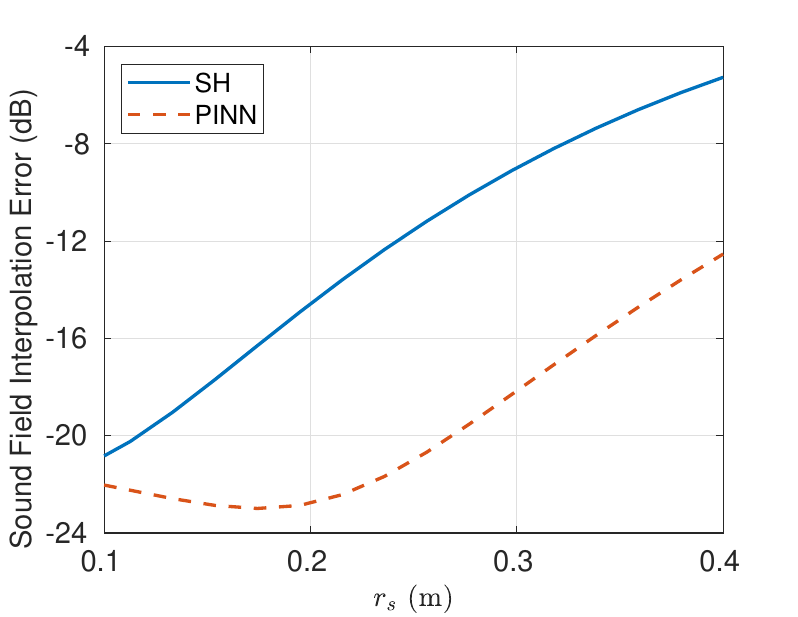}
\caption{Soundfield interpolation error using SH method and PINN method as a function of the sphere radius $r_s$.}
\label{fig:soundfield_error}
\end{figure}

We denote the soundfield interpolation error at radius $r$ as
\begin{IEEEeqnarray}{rcl}
    \epsilon_r = \frac{ \sum_{b=1}^{400} \left( p(n_b,\mathrm{x}_b,\mathrm{y}_b,\mathrm{z}_b)-\hat{p}(n_b,\mathrm{x}_b,\mathrm{y}_b,\mathrm{z}_b) \right)^2}{ \sum_{b=1}^{400} p(n_b,\mathrm{x}_b,\mathrm{y}_b,\mathrm{z}_b) ^2 }. \ \ \
\end{IEEEeqnarray}
Figure~\ref{fig:soundfield_error} shows the soundfield interpolation error of the two methods in dB.
For $r_s$ from $0.2$ m to $0.4$ m, the PINN method outperformed SH by approximately $8$ dB. 
The difference in interpolation error between the two methods is comparatively smaller for $r_s < 0.2$ m. Nevertheless, the PINN method consistently had a lower interpolation error.

\begin{figure}[t]
\centering
\includegraphics[trim={0.15cm 0cm 0.4cm 0.6cm},clip,height=4.4cm]{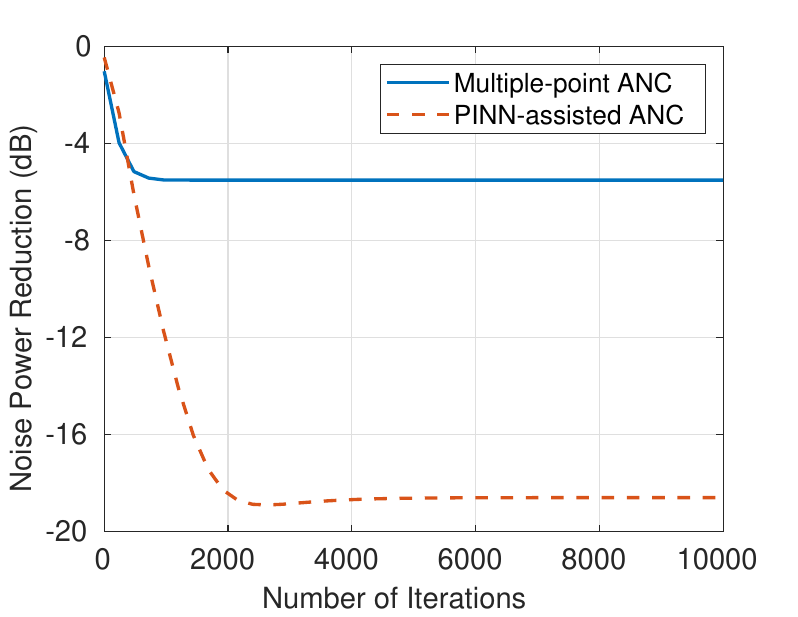}
\caption{Noise power reduction at the two ear locations achieved by
the multiple-point ANC and the PINN-assisted ANC system.}
\label{fig:ear_reduction}
\end{figure}

As our PINN method showed better soundfield interpolation than SH method in general, we omit evaluation of ANC system using SH-interpolated signals. To evaluate ANC performance, we define the noise reduction level at the two ear locations as
\begin{IEEEeqnarray}{rcl}
    \varepsilon = \frac{ \sum_{v=1}^2 \eV(n) ^2 }{ \sum_{v=1}^2 p_{v}(n) ^2 },
\end{IEEEeqnarray}
where $p_{v}(n)$ is the primary noise signal at the two ears. 
Figure~\ref{fig:ear_reduction} shows the noise reduction level achieved by the multiple-point and PINN-assisted ANC system. Both systems use the FxLMS algorithm with a step size of $\mu$ = 1$\times$10$^{-5}$ over $10000$ iterations.
While the initial noise power reduction rate in the first 500 iterations is similar for both ANC systems, our PINN approach achieved $-13$ dB more steady-state noise power reduction than the multiple-point ANC system.

\begin{figure}[!h]
    \centering
    \subfloat[]{
    \includegraphics[trim={0.1cm 0.1cm 0.8cm 0.5cm},clip,height=3.3cm]{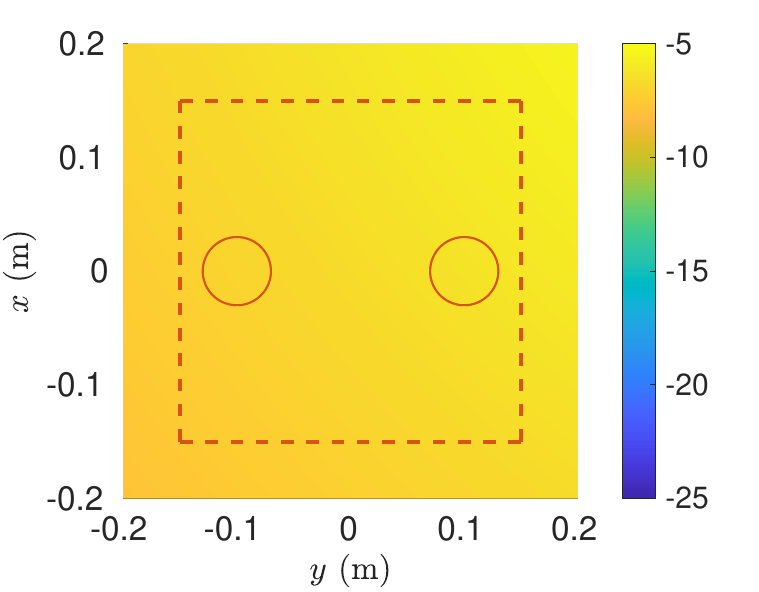}
    \label{fig:noise}
    }\\
    \subfloat[]{
    \includegraphics[trim={0.1cm 0.1cm 0.8cm 0.55cm},clip,height=3.3cm]{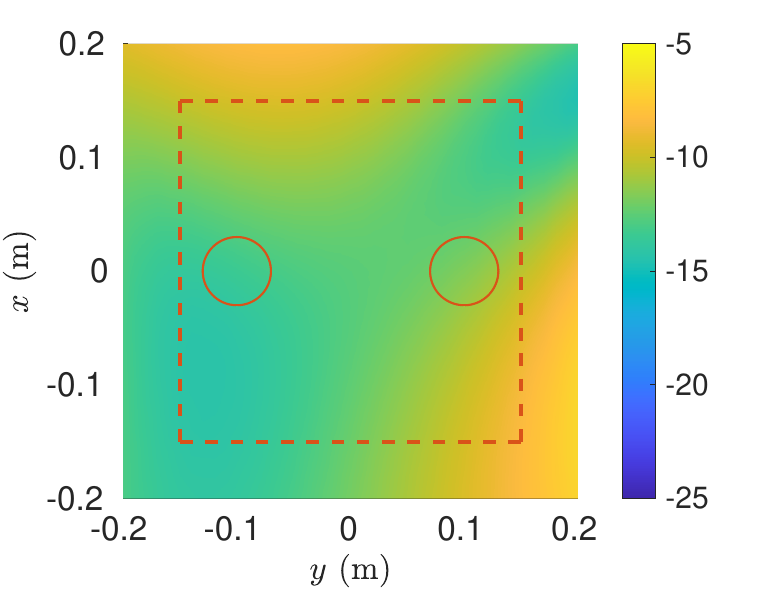}
    \label{fig:fxlms}
    }
    \hspace*{-0.1em}
    \subfloat[]{
    \includegraphics[trim={0.1cm 0.1cm 0.8cm 0.5cm},clip,height=3.3cm]{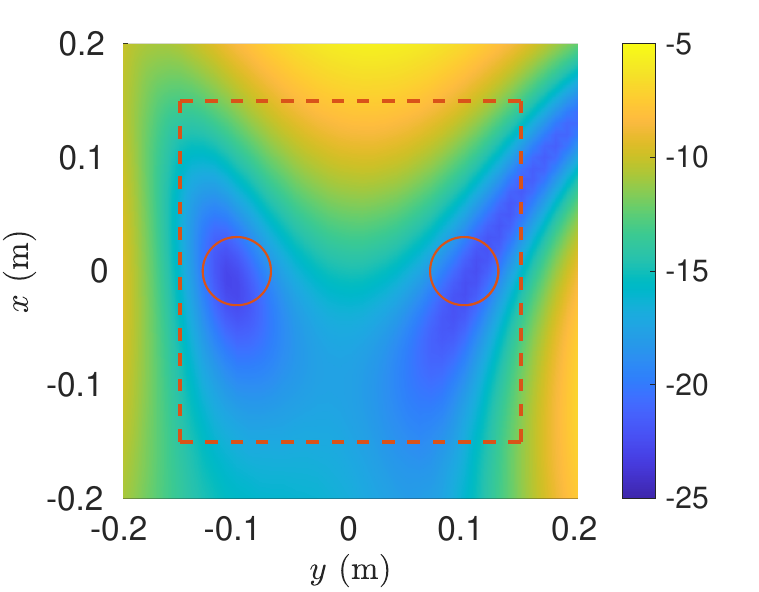}
    }
    \label{fig:pinn_fxlms}
    \caption{Signal power (in dB) in the $xy$-plane: (a) Primary noise field, and residual noise field for (b) multiple-point ANC system and (c) PINN-assisted ANC system. The dotted square is the projection of monitoring microphones on the $xy$-plane and the two circles depict two $r = 0.03$ m regions around the ears.}
    \label{fig:compare_3}
\end{figure}

In Figure~\ref{fig:compare_3}, we evaluated the signal power after convergence of the FxLMS algorithm in the $xy$-plane for $441$ evaluation points, which are evenly spaced from $-0.2$ m to $0.2$ m along $x$ and $y$ axes.
Figure~\ref{fig:compare_3} (a) is the original primary noise field.
Figure~\ref{fig:compare_3} (b) and (c) shows the residual noise power in multiple-point ANC system and PINN-assisted ANC system, respectively. In the dotted region, the PINN-assisted ANC showed an overall better noise reduction performance than the multiple-point ANC, and $-10$ dB lower residual noise field around the two ear regions.
This is due to the multiple-point ANC system reducing noise around monitoring microphones, whereas the PINN-assisted ANC system reduces noise around the virtual error microphone at the two ear locations.

\section{Conclusion}\label{sec:con}
In this paper, we proposed a practical ANC system with monitoring microphones placed outside the ROI. 
Using a PINN, the soundfield at the ROI is interpolated from the microphone signals, and the interpolated soundfield is used to reduce noise at the virtual microphone positions.
Our results show that the PINN method achieves an overall better interpolation result than SH method.
The noise attenuation performance of our PINN-assisted ANC system is also shown to exceed the multiple-point ANC system.

The proposed PINN model is simple and computationally efficient compared to deep neural networks with numerous parameters. 
We plan to quantitatively compare with other methods and evaluate PINN's computational advantage.
Additionally, we will extend our system to spatial ANC and replace the FxLMS algorithm with a machine learning-based controller, such as Deep MCANC~\cite{deliang2023} and DNoiseNet~\cite{DNoiseNet}.

\vfill\pagebreak


\bibliographystyle{IEEEbib}
\label{sec:refs}
\bibliography{refs.bib}

\end{document}